\documentclass[epj]{svjour}
\usepackage{graphics}
\usepackage{epsfig} 
\usepackage {array} 
\usepackage {amsmath} 
\usepackage {lscape}
\providecommand\bb{\boldsymbol{\rm b}}
\providecommand\bB{\boldsymbol{\rm B}} 
\providecommand\be{\boldsymbol{\rm e}} 
\providecommand\bE{\boldsymbol{\rm E}}

\providecommand\bj{\boldsymbol{\rm j}}

\providecommand\bn{\boldsymbol{\rm n}} 
 
\providecommand\bU{\boldsymbol{\rm U}}

\providecommand\bz{\boldsymbol{\rm z}} 
 
\providecommand\bmB{\overline{\boldsymbol{\rm B}}} 
 
\providecommand\bmU{\overline{\boldsymbol{\rm U}}} 
\providecommand\bmE{{\cal E}}
\providecommand\rm{R_m} 
\begin{document}
%
\title{Influence of inner and outer walls electromagnetic properties on the onset of a stationary dynamo.}
\author{R. Avalos-Zu\~niga\inst{1} \and F. Plunian\inst{1}
\thanks{\emph{email:} Franck.Plunian@hmg.inpg.fr}%
}                     
%
%
\institute{Laboratoires des Ecoulements G\'{e}ophysiques et Industriels, 
B.P. 53, 38041 Grenoble Cedex 9, France}
\date{Received: date / Revised version: date}
%
\abstract{To study the onset of a stationary dynamo in the presence of inner or outer walls of various electromagnetic properties, we propose a simple 1D-model in which the flow is replaced by an alpha effect. The equation of dispersion of the problem is derived analytically. It is solved numerically for 
walls of different thicknesses and of electric conductivity and magnetic permeability different from those of the fluid in motion. We also consider walls in the limit of infinite conductivity or permeability. 
} 
\authorrunning{R. Avalos-Zu\~niga \& F. Plunian}
\titlerunning{Influence of walls electromagnetic properties onto the onset of a stationary dynamo.}
\maketitle
\section{Introduction} 
\label{intro}
\begin{figure} 
  \begin{flushright} 
  \begin{tabular}{@{\hspace{0cm}}l@{\hspace{-0.7cm}}l@{\hspace{0cm}}} 
    \raisebox{5cm}{$\Gamma$} &
  \epsfig{file=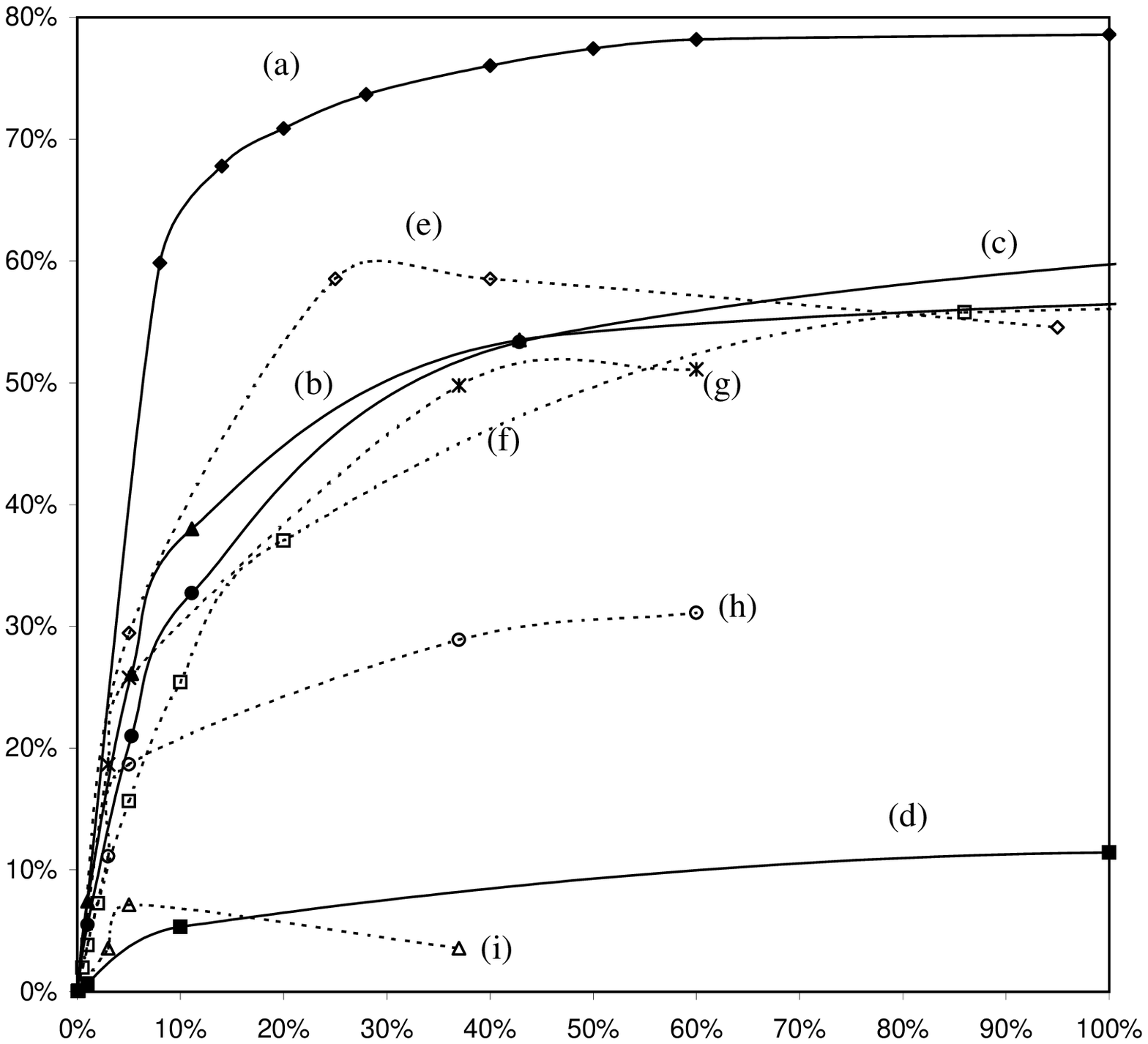,width=0.7\textwidth} 
    \\*[-1cm] 
     & \parbox{0.5\textwidth}{\hspace{6cm} $e/R$} 
    \end{tabular} 
\end{flushright} 
    \caption{Threshold reduction rate $\Gamma$ versus $e/R$ for (a) the Von Karman sodium experiment \cite{Ravelet05}, the Kumar and Roberts flow for (b) $R_m > 0$ and (c) $R_m < 0$ \cite{Sarson96}, (d) the Karlsruhe experiment \cite{Avalos03}, (e) the Perm experiment \cite{Frick02}, (f) the Riga experiment \cite{Avalos03}, the s1t1 flow with $\epsilon = $ (g) 0.35, (h) 0.3, (i) 0.25 \cite{Kaiser99}.}
    \label{fig:synthese}
   \end{figure}

A number of experimental devices have been built in the last years aiming at
producing dynamo action (for reviews see e.g. \cite{Gailitis02} and \cite{Radler02}). Such a device is generally made of a container in which some liquid
metal is put into motion.
In a previous study \cite{Avalos03} we considered the influence of the electromagnetic properties of the container outer wall onto the onset of dynamo action for
the Riga (Latvia) and Karlsruhe (Germany)  experiments. The results depend on which type of dynamo instability is obtained. For stationary solutions like in the Karlsruhe experiment \cite{Stieglitz01}\cite{Muller04}, the reduction of the dynamo instability threshold is monotonic versus the conductivity and the permeability of the outer wall. For oscillatory solutions like in the Riga experiment \cite{Gailitis00}\cite{Gailitis01}, there are additional eddy currents in the outer wall. These currents
produce an additional dissipation opposed to the reduction of the threshold. In that case, the reduction of the dynamo instability threshold versus the conductivity and the permeability of the outer wall is not monotonic anymore.\\
These results are consistent with other studies aiming at studying the influence
of the thickness of a stagnant outer layer conducting fluid (or equivalently of an outer wall with the same conductivity as the fluid)
on the dynamo threshold. Various inner flow geometries have been considered leading to either stationary \cite{Ravelet05}\cite{Sarson96} or oscillatory \cite{Kaiser99}\cite{Frick02} solutions.
In Fig. \ref{fig:synthese} we give a synthesis of these results. For that we plot the threshold reduction rate $\Gamma$ versus $e/R$ where $\Gamma = 1 - R_m(e/R)/R_m(0)$, $e$ is the thickness of the stagnant outer layer, $R$ the radius of the fluid container and $R_m$ the magnetic Reynolds number defined by $R_m = UR/\eta$ where $U$ is a characteristic flow intensity and $\eta$ the magnetic diffusivity. 
The stationary solutions correspond to the full curves (a-d) and are increasing monotonically versus $e/R$. The non-stationary solutions correspond to the dashed curves (e-i) and reach a maximum versus $e/R$ (though not obvious for the curves (f) and (h) it is actually the case). 
The influence of the electric conductivity of an inner core has also been investigated \cite{Schubert01}. Again the same conclusions as \cite{Avalos03} have been found: for stationary (resp. oscillating) solutions, the dynamo threshold decreases monotonically (resp. reaches a minimum) when increasing the conductivity of the inner core. \\
Given the high difficulties to build a dynamo experiment, reducing the threshold by changing the electromagnetic boundary conditions is of course of high interest for the experimenter. 
The influence of electromagnetic boundary conditions onto dynamo action has also been the object of different studies relevant to planets, and stars. In the case of Earth-like planets for example, the influence of a conducting solid inner core onto the dynamo action produced by the outer core fluid motion has been studied by different ways, using either a prescribed $\alpha^2$-effect
\cite{Schubert01}, a prescribed $\alpha$-effect and buoyancy \cite{Hollerbach95} or a direct resolution of the full convective dynamo model \cite{Wicht02}. The main issue of these studies was to identify whether the inner core has a stabilizing effect
on the reversals of the dipole component of the magnetic field. So far there is no definite answer to this problem as these different studies lead to contradictory results. In the case of Solar-like stars, it has been shown \cite{Thelen00} how some magnetic features (like the PDF of the magnetic field strength) observed at the surface of the star 
could give some indications on the relevant magnetic boundary conditions of a turbulent convective dynamo model.\\
Magnetic boundary conditions may also be important in the case of Fast Breeder Reactors (FBR). Indeed, though these industrial installations have not being designed to produce dynamo action, they share some common features with the Karlsruhe experiment. In the core of a FBR, the liquid sodium 
flows in an array of a large number of parallel straight tubes called
assemblies. In each of them the flow geometry is again composed of a periodic array of single helical vortices. One has shown \cite{Plunian99} the existence of an $\alpha$-effect similar to the Karlsruhe experiment in each assembly.
Though such an $\alpha$-effect is not sufficient to generate a dynamo instability for a core with homogeneous electromagnetic properties, the question remains when
the  walls of the assemblies are made, for example, of ferromagnetic steel (relative permeability of the order $10^3$ and relative conductivity of order 1) \cite{Soto99} and when the array of assemblies is surrounded by a belt of ferromagnetic material (as it is the case in the FBR Phenix). In that case both inner and outer walls electromagnetic properties may be important.
Indeed a reduction of the dynamo threshold leading to some dynamo instability within the core of the FBR could imped
the right working of the reactor and lead for example to an emergency breakdown.
\\

In the present paper we consider a simple 1D-model of a stationary dynamo in order to 
evaluate the relative importance of inner and outer wall electromagnetic properties onto
the onset of dynamo action. For that we consider two types of boundary conditions: either
an outer wall and isolating medium outside or periodic inner walls. In both cases we not only vary the relative conductivity and permeability of the walls but also their thicknesses. 

\section{A simple 1D-model}
\subsection{An anisotropic $\alpha$-effect}
\label{sec:alphaeffect}
  We use the kinematic approach consisting in solving the
induction equation for a given motion. This equation
reads  \begin{equation}     \frac{\partial\bB}{\partial t}=\nabla\times
(\bU\times\bB)+\eta \nabla^{2}\bB,\;\;\;\;\nabla.\bB = 0, 
\label{inductiondim} \end{equation}  where, again, $\eta$ means the magnetic
diffusivity of the fluid, $\bB$ the magnetic field and $\bU$ the fluid
velocity.\\ As we are not interested by a flow geometry in particular but only
by the influence of the boundary conditions onto the onset of dynamo action,
we assume that the interactions of the flow with the magnetic field
can be represented by an anisotropic $\alpha$-effect like in the Karlsruhe experiment \cite{Plunian02} \cite{Radler02b} or in the
core of a FBR \cite{Plunian99}. Following the lines of mean--field
dynamo theory \cite{Krause80} the magnetic field $\bB$ and the
fluid velocity $\bU$  are expressed as sums of mean fields, $\bmB$ and
$\bmU$,  and fluctuating fields, $\bB'$ and $\bU'$. Here the mean is defined
by a space average of the original field. Assuming $\bmU$ = 0, the mean part
of the induction equation  (\ref{inductiondim}) writes  \begin{equation}  
\frac{\partial \bmB}{\partial t} =     \nabla \times \bmE + \eta
\nabla^{2}\bmB,   \;\;\,\,\,\, \;\;\,\,\,\, \nabla.\bmB = 0,   
\label{inducmean}   \end{equation} where $\bmE$ is a mean electromotive force
due to the fluid motion  given by 
 \begin{equation}
 \bmE = \overline{\bU'
\times \bB'}.
 \end{equation}
We may consider $\bmE$ as a functional of
$\bU'$ and $\bmB$. 
Let us accept the assumption usually adopted in the
mean-field context that $\bmE$   in a given point in space and time depends on
$\bmB$ only via the components of  $\bmB$ and their first spatial derivatives
in this point.  This is reasonable for sufficiently small variations of
$\bmB$ in space and time.   For a first approximation, on which we restrict
ourselves here, we consider no other   contributions to $\bmE$ than that
describing the $\alpha$-effect,    that is, we ignore all contributions to
$\bmE$ containing derivatives of $\bmB$.   In addition, the $\alpha$-effect is
assumed to act in the $xy$-plane only, where $x$, $y$ and $z$ are the
Cartesian coordinates. This corresponds to a flow for example independent of
$z$.  Then (\ref{inducmean}) writes in the form \begin{equation}   
\frac{\partial \bmB}{\partial t} =-\nabla \times
[\alpha(\bmB - (\hat{\bz}.\bmB)\hat{\bz})] + \eta   \nabla^{2}\bmB,
\;\;\,\,\,\, \nabla.\bmB = 0   
\label{inducmean2}  
\end{equation}  
where $\alpha$ is a scalar quantity.
Such a model has proved to be sufficiently realistic for both cases, the Karlsruhe experiment \cite{Plunian02}\cite{Radler02b} and the FBR core \cite{Plunian99}. However for the latter case, there is an additional mean flow $\bmU$ along the $z$-direction which affects the model. The corresponding discussion is postponed to section \ref{sec:conclusion}.
\subsection{Model parameters} 
\begin{figure}
  \begin{tabular}{@{\hspace{-2.5cm}}l@{\hspace{0cm}}} 
  \epsfig{file=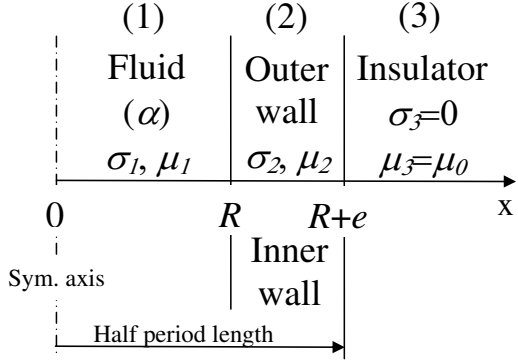,width=0.9\textwidth} 
    \\*[-15cm]  
    \end{tabular} 
    \caption{Scheme of the 3 regions of the non periodic problem (above) and of the 2 regions
    of the periodic problem (below).}   
  \label{fig:dessin}
 \end{figure}

We consider three regions $l$ (=1, 2 or 3) symmetric with respect to
the plane $x=0$ and infinite in the $y$ and $z$ directions. They are defined by their respective size along $x$ ($\pm x \le R$, $R \le \pm x \le R+e$ and $\pm x \ge R+e$),
conductivity ($\sigma_1$, $\sigma_2$, $\sigma_3$), permeability ($\mu_1$, $\mu_2$, $\mu_3$) and
$\alpha$-effect ($\alpha_1=\alpha$, $\alpha_2=0$, $\alpha_3=0$) where $\alpha$ is a steady scalar quantity which does not
depend on $x$, $y$ nor $z$.\\
We also consider two types of
boundary conditions in the $x$-direction (see Fig. \ref{fig:dessin}). The problem in which the region 3 is
an insulator ($\sigma_3=0$) and extends to infinity with $\bmB(x)\rightarrow  0$ when $x
\rightarrow\infty$ is called the non-periodic problem. In that case the region 2 corresponds to an outer wall.
 The problem in which the region 3 does not exists and
$\bmB(x=R+e)=\bmB(x=-R-e)$ is called the periodic problem. In that case the region 2 corresponds to periodic inner walls like the walls of the assemblies of a FBR.
\subsection{Reduction of the basic equations}
The solutions of (\ref{inducmean2}) can be represented as series of Fourier
modes proportional to $\exp(ijy+ikz)$.
As $\alpha$ does not depend neither on $y$ nor
$z$, each ($j,k$)-mode is independent from each other. 
As $\alpha$ is steady we may expect solutions
varying like $\exp(pt)$ in time, with the real part of $p$ being the growth
rate of the magnetic field. 
In the rest
of the paper we consider only the mode $j=0$, for sake of simplicity.
Then, for a given $k$, we may look for $\bmB$ in the form 
\begin{equation}
\bmB = \Re [\bb(x) \exp(pt+ikz)] \;\;\; \mbox{with} \;\;\; \bb = (-ika,
b,\partial_{x}a), \label{bb} 
\end{equation}
$a$ and $b$ being functions of $x$ only.
Replacing (\ref{bb}) in
(\ref{inducmean2}), we find the following equations 
\begin{eqnarray}
(\eta_l \Delta-p)a_l-\alpha_lb_l=0 \label{eq:equations1}\\
(\eta_l \Delta-p)b_l-\alpha_lk^2a_l=0 \label{eq:equations2}
\end{eqnarray}
where $\eta_l=1/\sigma_l\mu_l$, $\Delta=\partial_{x^2}-k^2$
and $l=1,2,3$ ($l=1,2$) for the non-periodic (periodic) problem.\\
We can show that there exists two sets of independent solutions depending on
the parity of $a_1(x)$ and $b_1(x)$. Indeed, from (\ref{eq:equations1}) and
(\ref{eq:equations2}) $a_1$ is solution of $L(a_1)=0$ with $L=(\eta_1
\Delta-p)^2-\alpha^2k^2$. This operator $L$ is linear and leaves the
parity of the function unchanged. Therefore writing $a_1$ as the sum of an odd
and even functions $a_1=a_1^o+a_1^e$ we find that $L(a_1^o)=L(a_1^e)=0$.
Consequently, $a_1^o$ and $a_1^e$ are two independent solutions of $L$. 
From (\ref{eq:equations1}) and (\ref{eq:equations2}), it is
easy to show that $b_1$ has the same parity as $a_1$.
Therefore it is sufficient to solve (\ref{eq:equations1}) and
(\ref{eq:equations2}) for each parity, the general solution being a linear combination of them.
The solution at $x=0$ is given by
 $\bb=(0,0,b_z)$ if $a_1(x)$ is odd and $\bb=(0,b_y,b_z)$ if $a_1(x)$ is even.
\section{Method of resolution}
\subsection{General solutions} 
The general form of the solutions of equations (\ref{eq:equations1}) and
(\ref{eq:equations2}) write: 
\begin{eqnarray}
a_1&=&\frac{1}{k}(A_{1c}^+ \cosh\omega_1^+x + A_{1c}^- \cosh\omega_1^-x \nonumber \\
&& +\;\; A_{1s}^+ \sinh\omega_1^+x + A_{1s}^- \sinh\omega_1^-x) \nonumber \\ 
b_1&=&B_{1c}^+ \cosh\omega_1^+x + B_{1c}^- \cosh\omega_1^-x \nonumber \\ 
&+& B_{1s}^+ \sinh\omega_1^+x + B_{1s}^- \sinh\omega_1^-x \label{apoddsol}\\ 
a_2&=&\frac{1}{k}(A_{2c}\cosh\omega_2x + A_{2s}\sinh\omega_2x) \nonumber\\ 
b_2&=&B_{2c} \cosh\omega_2x + B_{2s}\sinh\omega_2x\nonumber \\     
a_3&=&\frac{A_3}{k}\exp(-\omega_3x)\nonumber \\
b_3&=&B_3\exp(-\omega_3x)\nonumber
\end{eqnarray}
with
\begin{eqnarray}  
(\omega_1^{\pm})^2&=&k^2+\frac{p \pm \alpha k}{\eta_1},\;\;
(\omega_2)^2=k^2+\frac{p}{\eta_2},\label{omega123}\\
(\omega_3)^2&=&k^2+\frac{p}{\eta_3}\;\;\; \mbox{and} \;\;\;\Re(\omega_3)>0 \nonumber
\end{eqnarray}
and where $a_3$ and $b_3$ are the solutions in the region 3 for the non
periodic problem only. In that case we applied the condition $\bb_3
\rightarrow 0$ when $r \rightarrow \infty$. Furthermore, for sake of generality, we shall replace $\sigma_3$ by 0 only in the numerical applications.\\
Applying the appropriate symmetry conditions in $x=0$, the following relations
are found for
the even solutions:
\begin{equation}
A_{1s}^+=A_{1s}^-=B_{1s}^+=B_{1s}^-=0.
\label{eq:even1}
\end{equation}
From (\ref{eq:equations1}) we have the additional relations:
\begin{equation}
 A_{1c}^+=B_{1c}^+,\;\;\;\;\;A_{1c}^-=-B_{1c}^-
\label{eq:even2}
\end{equation}
for both problems non-periodic and periodic.
For the even solutions of the periodic problem
the boundary condition $\bb_2(R+2e)=\bb_2(R) $ leads to the
additional relations  \begin{eqnarray}
A_{2c}\sinh\omega_2(R+e)&=&-A_{2s}\cosh\omega_2(R+e)\nonumber \\
B_{2c}\sinh\omega_2(R+e)&=&-B_{2s}\cosh\omega_2(R+e).
\label{eq:bceven}
\end{eqnarray}
For the odd solutions the same relations as (\ref{eq:even1}) and (\ref{eq:even2}) are found
with inverted subscripts $s$ and $c$. For the odd solutions of the periodic problem
the boundary conditions $\bb_2(R+2e)=-\bb_2(R) $ leads to the
additional relations given by (\ref{eq:bceven}) with, again, inverted subscripts $s$ and $c$.
\subsection{Boundary conditions} 
The normal component of $\bmB$, the tangential component of $\bmB/\mu$ and the
$z$-component of the electric field $E_z=\eta (\nabla \times \bmB)_z$ are
continuous across each interface $x=R$ and $x=R+e$. We can show that this set
of relations is sufficient to describe all the boundary conditions of the
problem. They write for $l$=1, 2 ($l=1$) corresponding to the non-periodic (periodic) case:   
\begin{eqnarray}  
a_{l}(x=x_l)&=&a_{l+1}(x=x_{l}) \nonumber\\    
\frac{1}{\mu_l}b_{l}(x=x_l)&=&\frac{1}{\mu_{l+1}}b_{l+1}(x=x_l)\nonumber\\
\frac{1}{\mu_l}a'_l(x=x_l)&=&\frac{1}{\mu_{l+1}}a'_{l+1}(x=x_l)\nonumber\\
\eta_lb'_l(x=x_l)&=&\eta_{l+1}b'_{l+1}(x=x_l)
\label{apBC}
\end{eqnarray}
where the prime denotes the $x$-derivative, $x_1=R$ and $x_2=R+e$.
\subsection{Resolution} 
Replacing (\ref{apoddsol}) into (\ref{apBC}) and applying (\ref{eq:even1}), (\ref{eq:even2}) and (\ref{eq:bceven}), we find a
system to solve.
The solution is non trivial only if the
determinant of the system is equal to zero. 
This determinant writes
\begin{equation}
1+\frac{M+S}{2}(X_1^+ +X_1^-) + MSX_1^+ X_1^- =0
\label{ap:carac}
\end{equation}
with $X_1^{\pm}=R\omega_1^{\pm}\tanh R\omega_1^{\pm}$ for the even solutions and
$X_1^{\pm}=R\omega_1^{\pm}\coth R\omega_1^{\pm}$ for
the odd solutions.\\
For both even and odd solutions of the non-periodic problem,
$M$ and $S$ are defined by
\begin{equation}
M=\frac{m}{R\omega_2}\frac{1+\frac{n\omega_3}{\omega_2}\tanh e\omega_2}{\frac{n\omega_3}{\omega_2}+\tanh e\omega_2}\;,\;\;
S=\frac{s}{R\omega_2}\frac{1+\frac{r\omega_3}{\omega_2}\tanh e\omega_2}{\frac{r\omega_3}{\omega_2}+\tanh e\omega_2}.
\label{MSdefinition}
\end{equation}
For the periodic problem, $M$ and $S$ are defined by
\begin{eqnarray}
(M,S)&=&(m,s)\frac{\coth e\omega_2}{R\omega_2}\;\;\;\;\;\; \mbox{for
even solutions,}\label{pereven}\\
(M,S)&=&(m,s)\frac{\tanh e\omega_2}{R\omega_2}\;\;\;\;\;\; \mbox{for
odd solutions,}
\end{eqnarray}
with 
\begin{equation}
m=\frac{\mu_2}{\mu_1},\;\;\;
n=\frac{\mu_2}{\mu_3},\;\;\;s=\frac{\sigma_2}{\sigma_1},\;\;\;r=\frac{\sigma_2}{\sigma_3}.
\label{param}
\end{equation}

\section{Results}
\subsection{General remarks}
We can show that the
growth rate $p$ has no imaginary part as shown in Appendix A. An additional and simple
argument is that, as the $\alpha$-effect does not depend on $z$, there is no preferred
way along the $z$-direction for a magnetic wave to travel. Therefore, the
marginal instability solution corresponds to $p=0$. This leads to
$\omega_2=\omega_3=k$.\\
The results are given in terms of the dimensionless
quantities $m$, $n$, $s$, $r$ defined in (\ref{param}) and of $\hat{k}=kR$,
$\hat{p}=pR^2/\eta_1$, $\hat{\omega_l}=R\omega_l$, $\hat{e}=e/R$ and
$R_\alpha=R\alpha/\eta_1$. The hat is dropped in the rest of the paper
for sake of clarity.
Then for a given set of the parameters $m, n, s, r, k, e$ we seek solutions such that $p=0$ and $R_\alpha$ is
minimum.\\
From (\ref{omega123}) we see that replacing $R_{\alpha}$ by -$R_{\alpha}$ is
the same as replacing $k$ by $-k$ and also the same as replacing $\omega_1^+$
by $\omega_1^-$. As (\ref{ap:carac}) is
symmetric in $\omega_1^+$ and $\omega_1^-$, it is then sufficient to
consider only positive values of $k$ and $R_{\alpha}$.\\
For the non-periodic problem we set $\sigma_3=0$ (the region 3 being insulating).  This
corresponds to the limit $r\rightarrow\infty$ . It leaves (\ref{ap:carac}) unchanged but $\omega_3=k$ and
$S=s\frac{\tanh\omega_2e}{\omega_2}$.
\subsection{Influence of wall thickness}
\label{sec:e}
\begin{figure} 
  \begin{flushright} 
  \begin{tabular}{@{\hspace{0cm}}l@{\hspace{0cm}}l@{\hspace{0cm}}} 
    \raisebox{5cm}{$\log_{10}(R_{\alpha})$}&
  \epsfig{file=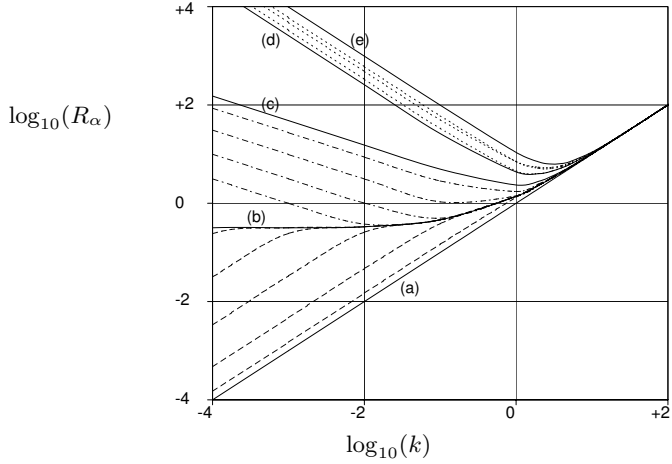,width=0.6\textwidth} 
    \\*[-1cm] 
     & \parbox{0.5\textwidth}{\hspace{3cm} $\log_{10}(k)$} 
    \end{tabular} 
\end{flushright} 
    \caption{The marginal curves $\log_{10}(R_{\alpha})$ versus $\log_{10}(k)$
for $m=n=1$, $s=10$ and different values of $e$. The dashed
curves between (a) and (b) correspond to the periodic even
solutions. From (a) to (b): $e$=0 (curve (a)),$e=1,10,10^2,10^3,10^4$ and 
$e\rightarrow \infty$ (curve (b)). The dotted-dashed curves between (b) and (c)
correspond to the non-periodic even solutions. 
From (c) to (b): $e$=0 (curve (c)),$e=0.1,1,10,10^2$ and 
$e\rightarrow \infty$ (curve (b)).
The dotted curves between (d) and (e)
correspond to the odd solutions for both problems (periodic and
non-periodic). They are obtained respectively in
the limit of large $e$ (curve (d)) and for $e=0$ (curve (e)).}    
\label{fig:bilan}    
\end{figure}

\begin{figure} 
  \begin{flushright} 
  \begin{tabular}{@{\hspace{0cm}}l@{\hspace{0cm}}c@{\hspace{0cm}}} 
    \raisebox{5cm}{}& 
  \epsfig{file=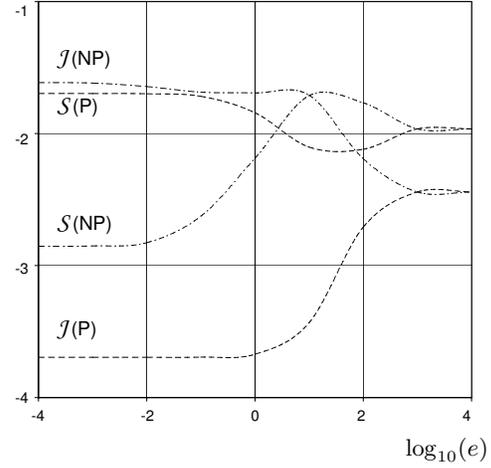,width=0.6\textwidth} 
    \\*[-1.cm] 
     & \parbox{0.5\textwidth}{\hspace{5cm} $\log_{10}(e)$} 
    \end{tabular} 
\end{flushright} 
    \caption{Dissipation rate $\cal{J}$ and work of the Lorentz forces rate $\cal{S}$ versus $\log_{10}(e)$ for $m=n=1$,
$s=10$ and $k=0.01$ for both problems periodic (P) and non-periodic (NP).}   
  \label{fig:dissipation}   
 \end{figure}

\begin{figure} 
  \begin{tabular}{@{\hspace{-1cm}}c@{\hspace{-3.16cm}}c@{\hspace{-3.16cm}}c@{\hspace{-3.16cm}}c@{\hspace{0cm}}} 
  \epsfig{file=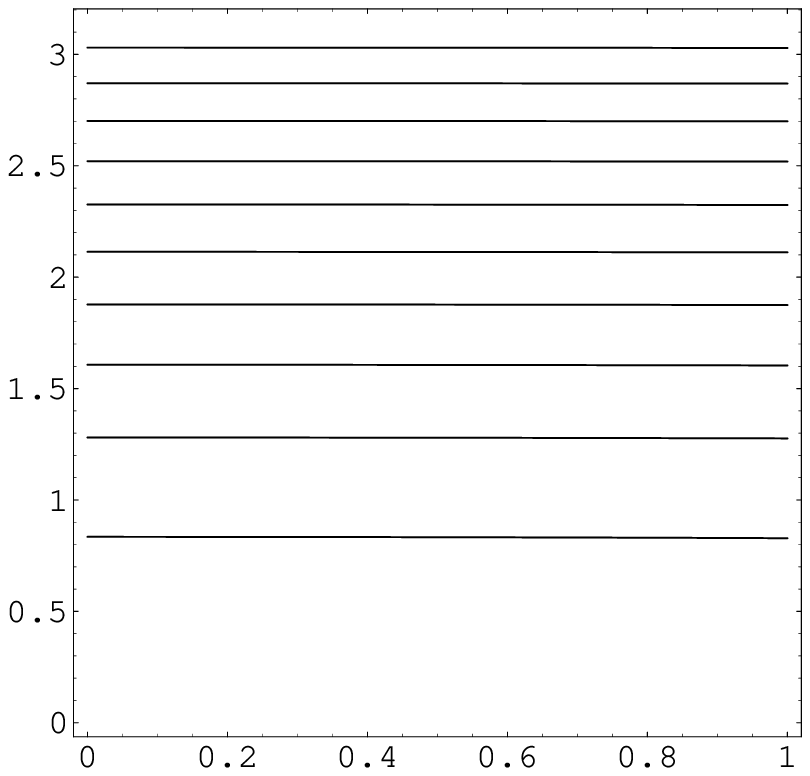,width=0.3\textwidth}& 
  \epsfig{file=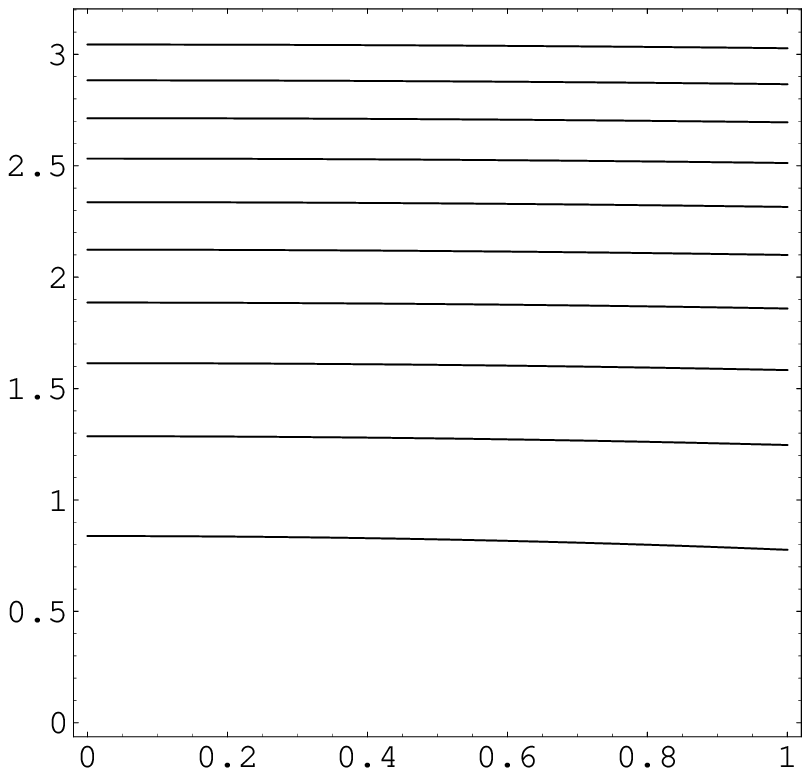,width=0.3\textwidth} &
  \epsfig{file=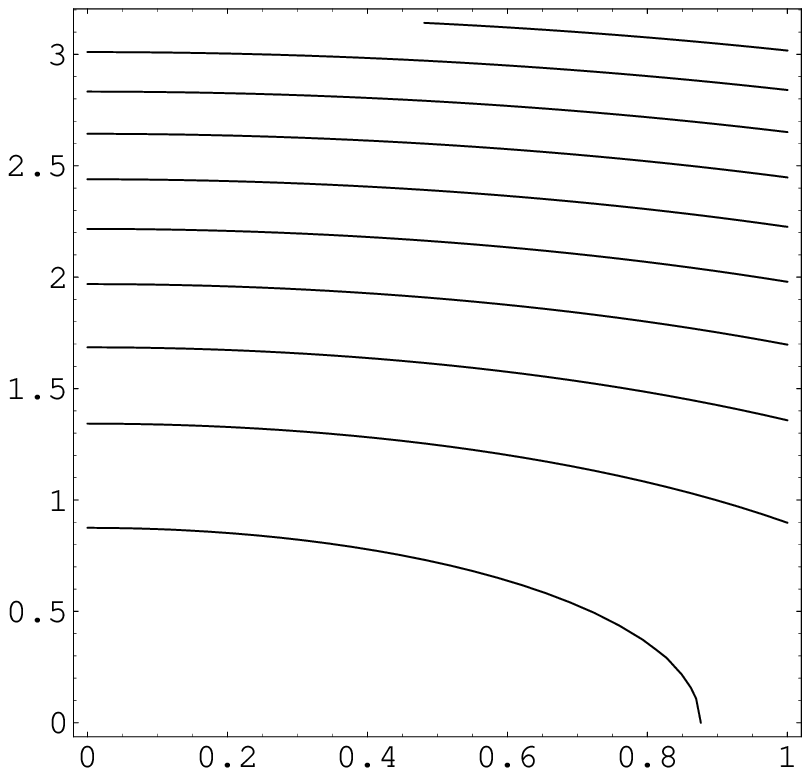,width=0.3\textwidth} &
  \epsfig{file=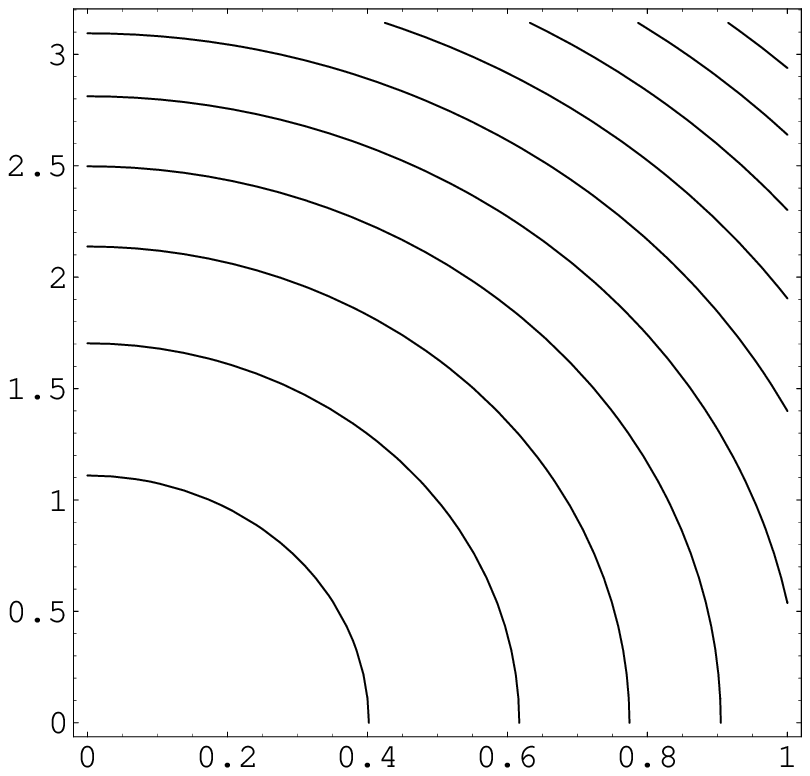,width=0.3\textwidth}
    \\ 
    {\hspace{-1.5cm}(a)} & {\hspace{-1.5cm}(b)} & {\hspace{-1.5cm}(c)} & {\hspace{-1.5cm}(d)}
    \\*[-5.4cm]
  \epsfig{file=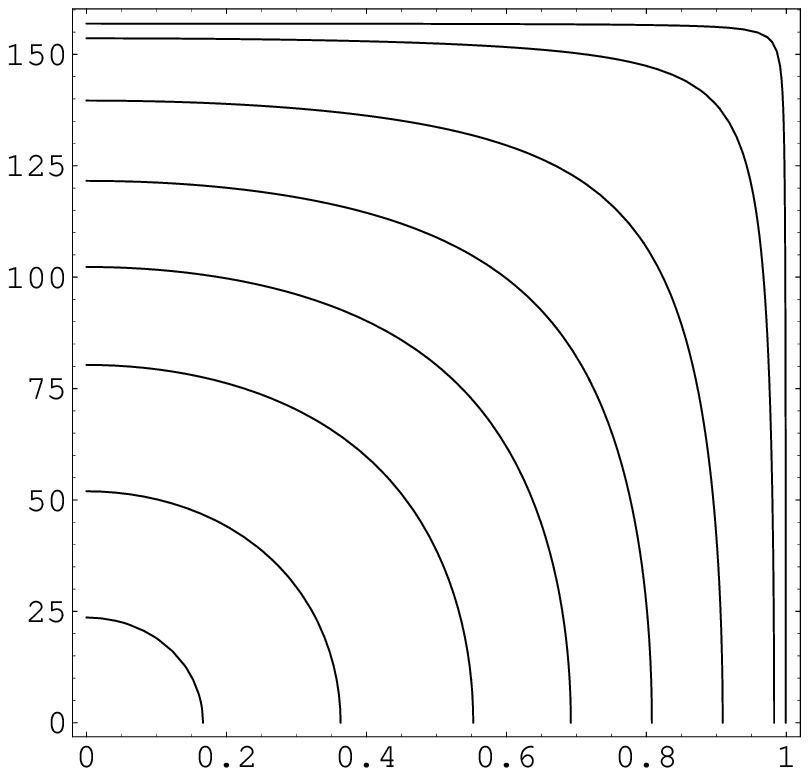,width=0.3\textwidth}& 
  \epsfig{file=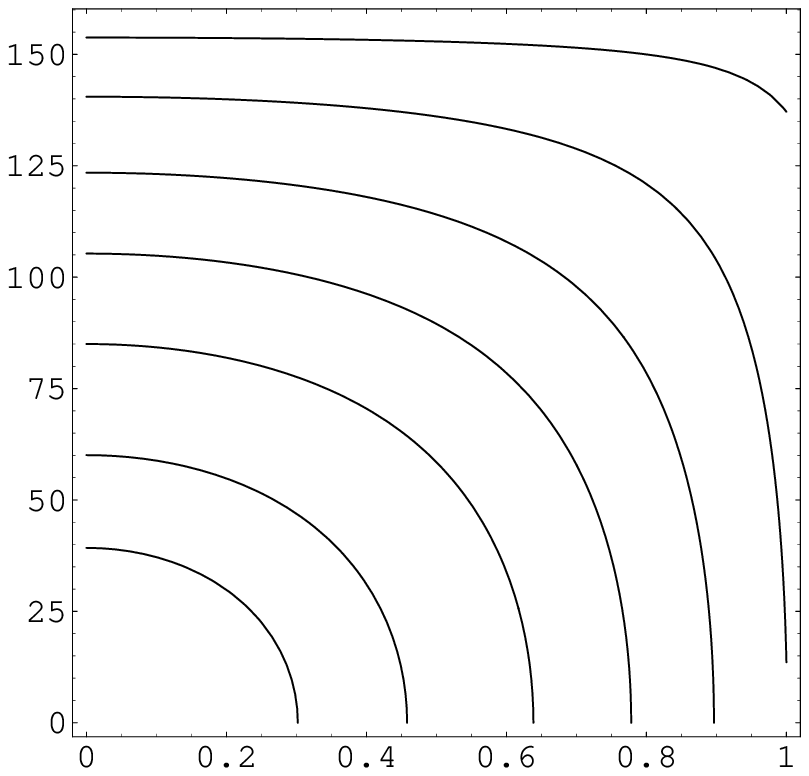,width=0.3\textwidth} &
  \epsfig{file=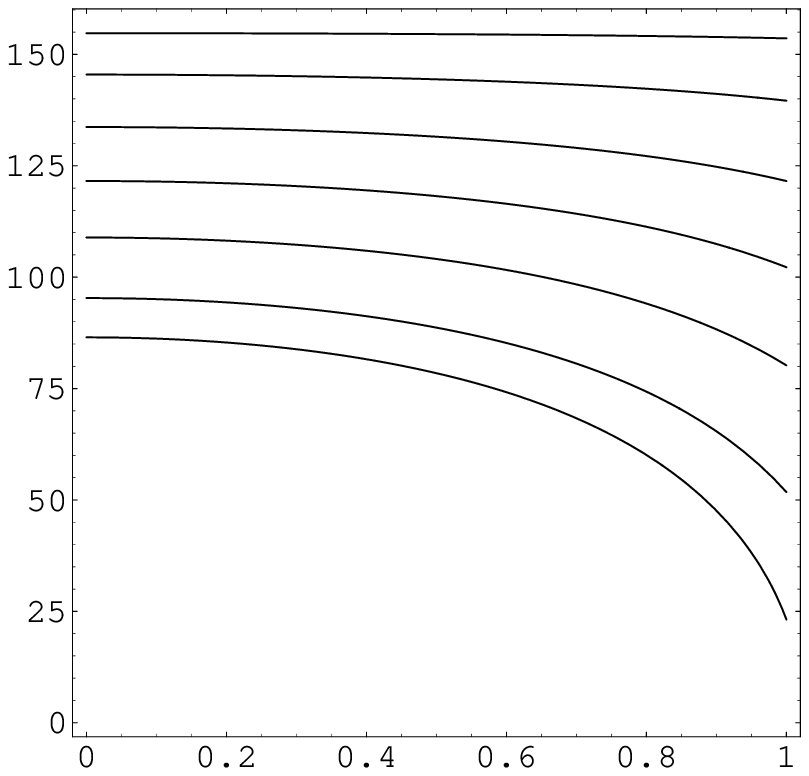,width=0.3\textwidth} &
  \epsfig{file=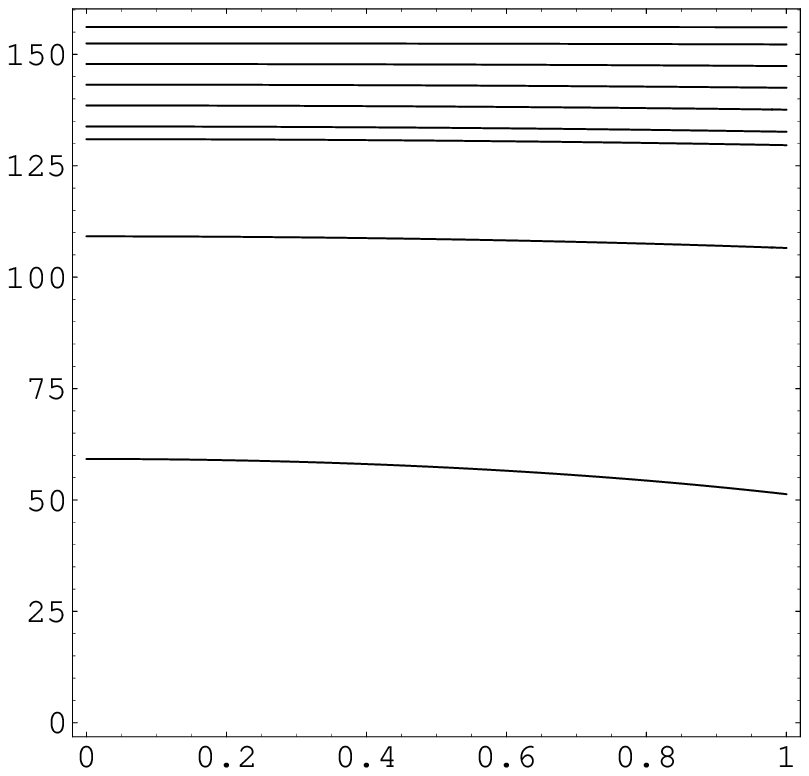,width=0.3\textwidth}
    \\ 
    {\hspace{-1.5cm}(e)} & {\hspace{-1.5cm}(f)} & {\hspace{-1.5cm}(g)} & {\hspace{-1.5cm}(h)}
    \end{tabular} 
    \caption{Isolines of $\Re[b(x)\exp(ikz)]$ in the $(x,z)$-plane ($0\le x\le1$) for $m=n=1$,
$s=10$, $k=0.01$, for both periodic ($0\le kz\le \pi / 100$) and non-periodic ($0\le kz\le \pi$) problems
    and different values of $e$. The isolines of the periodic problem are given for (a) $e=10^{-1}$,
    (b) $e=1$, (c) $e=10$ and (d) $e=10^2$.
    The isolines of the non-periodic problem are given for (e) $e=0$,
    (f) $e=10^{-2}$, (g) $e=10^{-1}$ and (h) $e=1$.}
    \label{fig:contours}
    \end{figure}

In Fig. \ref{fig:bilan} the marginal curves $\log_{10}(R_{\alpha})$ versus
$\log_{10}(k)$ are given for $s=10$, $m=n=1$ and different values of $e$. The
marginal curves of the odd solutions are located between the
curves (d) and (e) and therefore always above the marginal curves
of the even solutions located between the curves (a) and (c).
Therefore only the even solutions are present at the onset of the
dynamo action unless $k$ is large. The same comments apply when varying
$s$, $m$ or $n$. Therefore in the subsequent subsections we shall focus only on the even solutions.\\
The marginal curves
of the non-periodic problem are located
between the curves (c) and (b). They are always above those of the periodic
problem located between the curves (a) and (b). Then the periodic
problem appears to be always more unstable than the non-periodic problem.
However in the limit of large $ke$ both problems have the same marginal curve
(b). 
Indeed taking $ke>>1$ in (\ref{MSdefinition}) and (\ref{pereven}) leads to $(M,S)=(m,s)/k$ for both problems.
Also taking $ke>>1$ means that
the fluid is embedded between walls of infinite thickness (compared to the
vertical wave length of the field). Then no distinction can be found
between both problems. From asymptotic estimates, we can show that  
\begin{equation}  \lim_{ke \rightarrow \infty, k \rightarrow 0}
R_{\alpha}=(ms)^{-1/2} 
\end{equation} 
which corresponds to the asymptotic left
part of the curve (b).\\
The curve (a) is obtained for $e=0$ for the periodic problem (even solutions).
We can show that it is given by  
\begin{equation}
R_{\alpha}=k.
\end{equation}
which has already been obtained in other periodic problems
(without walls) in which an anisotropic $\alpha$-effect is the dynamo mechanism
(see e.g. \cite{Plunian99}\cite{Plunian02}).\\
The curve (c) is obtained for $e=0$ for the non-periodic problem (even solutions).
From asymptotic estimate we can show that for $k<<1$ it is
given by 
\begin{equation} 
R_{\alpha}=\sqrt{3n/mk}.
\end{equation}
In the limit
of small or large values of both $k$ and $ke$, asymptotic expansions in $k$
of $R_{\alpha}$ have been calculated. A synthesis
of these expansions is given in table \ref{tab:asymptotics}.
\begin{table*}  
  \begin{center}   
  \begin{tabular}{|l c|c|c|} \hline  
        &$R_{\alpha}$& $k<<1$ and $ke<<1$ & $k>>1$ and $ke>>1$\\ \hline
even &n.p.&
$\sqrt{\frac{n/m}{se+1/3}}\frac{1}{\sqrt{k}}+\frac{1}{\sqrt{k}}O(1)$&
$k+(\frac{\pi}{2})^2\frac{1}{k}-$ \\ \cline{2-3}
&per.& $\sqrt{(1+e/m)(1+e/s)}\;k+O(k)$ &
$(\frac{\pi}{2})^2
\sqrt{2}\frac{m+s+2\sqrt{2}ms}{m+s+\sqrt{2}}\frac{1}{k^2}+O(\frac{1}{k^2})$
\\
\hline odd && $\frac{c^2}{k}+\frac{1}{k}O(1)$&
$k+\pi^2(\frac{1}{k}-
\sqrt{2}\frac{m+s+2\sqrt{2}ms}{m+s+\sqrt{2}}\frac{1}{k^2})+O(\frac{1}{k^2})$
\\ \hline   
\end{tabular}  
  \end{center}  
  \caption{Asymptotic expansion in $k$ of $R_\alpha$ for both problems 
non-periodic and periodic. The parameter $c$
($\frac{\pi}{2}<c<\pi$) is solution of the equation  $\frac{1}{2se}(\tan
c+\tanh c)=-c$ for the non-periodic problem and $(\frac{m+s}{2}+msec\cot
c)(\frac{m+s}{2}+msec\coth c)=(\frac{m-s}{2})^2$ for the  periodic problem.}   
\label{tab:asymptotics}   
\end{table*}   
These expansions
fit very well to the slopes of the curves of Fig. \ref{fig:bilan}.\\
In the limit of small $ke$ and large $e$ these expansions also give $R_{\alpha}=O(e^{-1/2})$ for the
non-periodic and $R_{\alpha}=O(e)$ for the
periodic even solutions. Then for
increasing $e$, the non-periodic problem gets more unstable whereas the
periodic problem gets more stable. To get a qualitative explanation to this
striking difference between both problems, we sketch in Fig.
\ref{fig:dissipation} the dimensionless dissipation rate $\cal{J}$ 
(ratio of the Joule
dissipation to the magnetic energy, see details in Appendix B) versus
$e$. We also sketch on the same figure the dimensionless rate of the work of the Lorentz forces $\cal{S}$. Then
$R_{\alpha}$ is defined by $R_{\alpha}=\cal{J}/\cal{S}$. 
What is remarkable is that the decrease (increase) of $R_{\alpha}$ for the non-periodic (periodic) problem is not only due to a decrease (increase) of 
$\cal{J}$ but also to an increase (decrease) of $\cal{S}$. We also find that the dissipation in the wall in both cases is always negligible compared to the dissipation in the fluid. This shows that changing $e$ leads to a 
pure geometrical effect on the field and current lines, a consequence of it being the change of the dissipation in the fluid and work of the Lorentz forces. As an illustration we sketch in Fig.\ref{fig:contours} the isolines of $\Re[b(x)\exp(ikz)]$ for different values of $e$. These isolines correspond to the current density lines in the $(x,z)$-plane and also to the isolines of the $y$-component of the magnetic field. We see that increasing the thickness (from left to right) for both problems periodic (top) and non-periodic (bottom) changes indeed the geometry of these isolines.
Increasing $e$ leads to a bending (flattening) of the current lines for the periodic (non-periodic) problem, enhancing (decreasing) Joule dissipation, consistently with Fig.\ref{fig:dissipation}.
\subsection{Influence of wall conductivity and permeability}
\label{sec:cp}
\begin{figure} 
  \begin{flushright} 
  \begin{tabular}{@{\hspace{0cm}}l@{\hspace{0cm}}c@{\hspace{0cm}}} 
    \raisebox{5cm}{$R_{\alpha}$}& 
  \epsfig{file=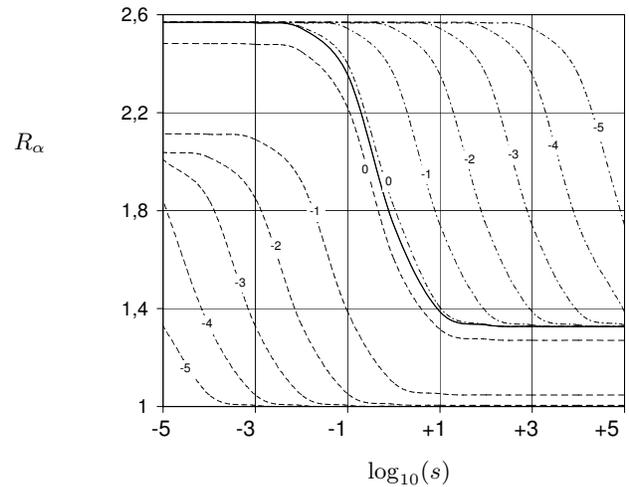,width=0.63\textwidth} 
    \\*[-1cm] 
     & \parbox{0.5\textwidth}{\hspace{3cm} $\log_{10}(s)$} 
    \end{tabular} 
\end{flushright} 
    \caption{The marginal curves $R_{\alpha}$ versus $\log_{10}(s)$
for  $k=m=n=1$ and different values of $e$. The dashed (dotted-dashed)
curves correspond to the periodic (non-periodic) even solutions.
The labels indicate $\log_{10}(e)$. The full line is common to both problems in
the limit of large $e$.}     \label{fig:sigma}    \end{figure}

\begin{figure} 
  \begin{flushright} 
  \begin{tabular}{@{\hspace{0cm}}l@{\hspace{0cm}}c@{\hspace{0cm}}} 
    \raisebox{5cm}{$R_{\alpha}$}& 
  \epsfig{file=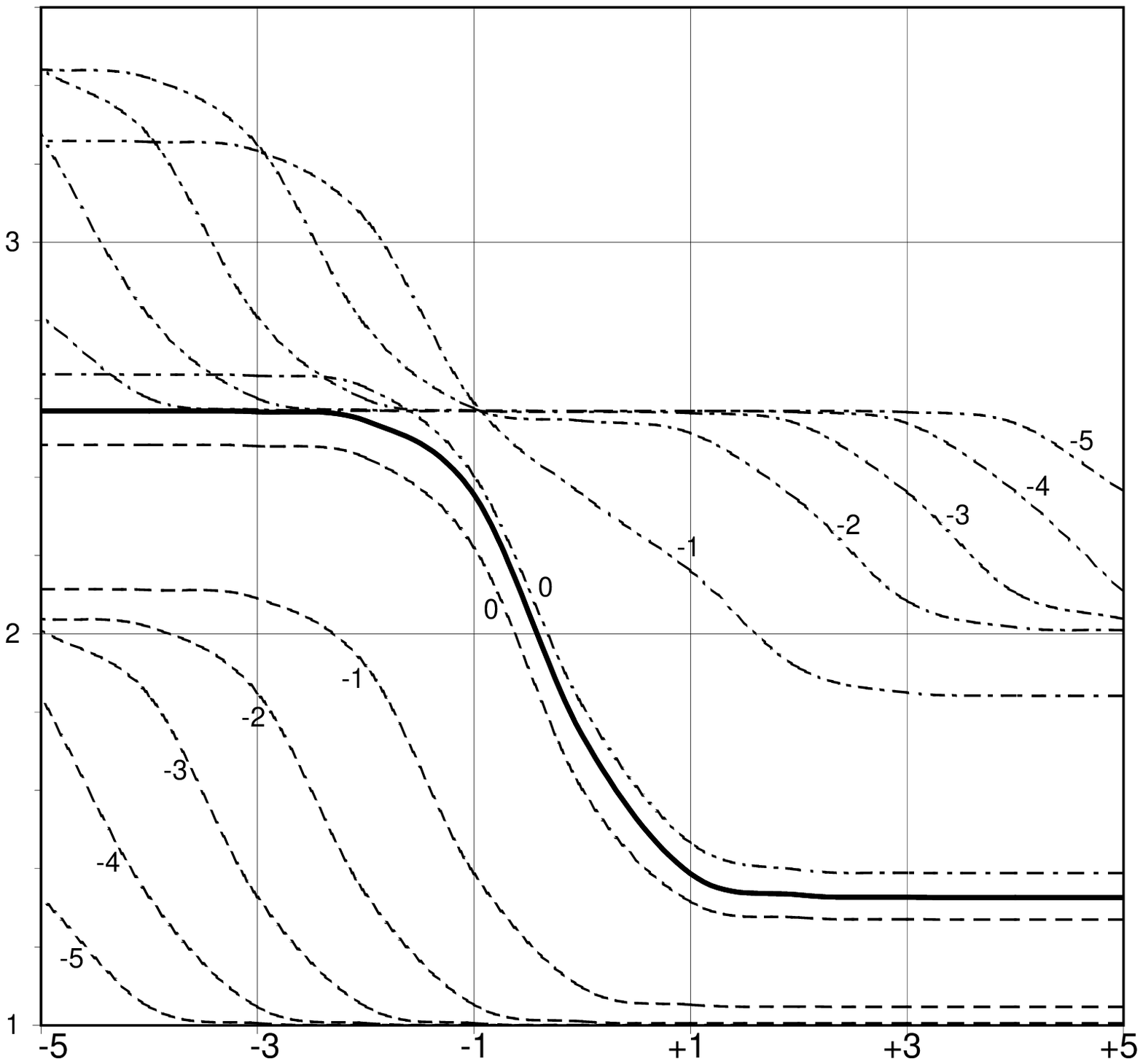,width=0.63\textwidth} 
    \\*[-1cm] 
     & \parbox{0.5\textwidth}{\hspace{3cm} $\log_{10}(m)$} 
    \end{tabular} 
\end{flushright} 
    \caption{The marginal curves $R_{\alpha}$ versus $\log_{10}(m)$
for  $k=s=1$, $n=m$ and different values of $e$. The dashed (dotted-dashed)
curves correspond to the periodic (non-periodic) even solutions.
The labels indicate $\log_{10}(e)$. The full line is common to both problems in
the limit of large $e$.}     \label{fig:mu}    \end{figure}

\begin{figure} 
  \begin{flushright} 
  \begin{tabular}{@{\hspace{0cm}}l@{\hspace{0cm}}c@{\hspace{0cm}}} 
    \raisebox{5cm}{$\log_{10}(m)$}&
  \epsfig{file=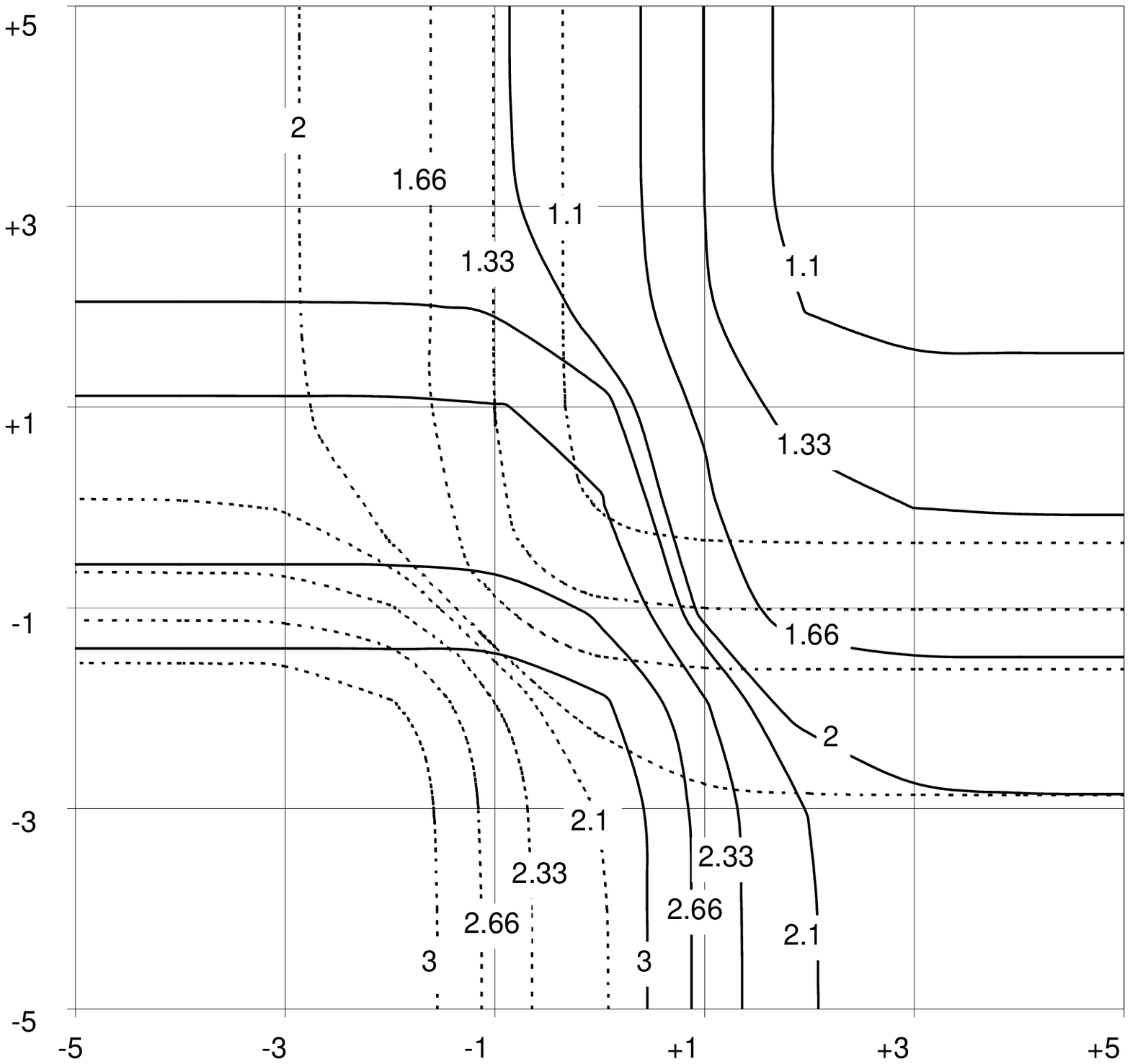,width=0.61\textwidth} 
    \\*[-1cm] 
     & \parbox{0.5\textwidth}{\hspace{3cm} $\log_{10}(s)$} 
    \end{tabular} 
\end{flushright} 
    \caption{Isolines of
$R_{\alpha}$ in the ($\log_{10}(s)$, $\log_{10}(m)$)-plane
for $k=1$, $e=0.1$ and $n=m$. The dashed (full)
curves correspond to the periodic (non-periodic) even solutions.}
    \label{fig:perm}
   \end{figure}

The expansions of table \ref{tab:asymptotics} for small values of both $k$ and
$ke$ also give the dependency of $R_{\alpha}$ versus $m$ and $s$. For large (resp.
small) value of $m$ and $s$, $R_{\alpha}=O((ms)^{-1/2})$ for the
non-periodic (resp. periodic) even solutions. This means that the
dynamo onset is easier to reach with walls of higher permeability and/or
conductivity. This is true also outside the previous asymptotic limits 
as depicted in Fig. \ref{fig:sigma}, Fig. \ref{fig:mu} and Fig. \ref{fig:perm}.\\ 
  
In Fig. \ref{fig:sigma} the
marginal curves $R_{\alpha}$ versus $\log_{10}(s)$ are given for $k=1$, $m=n=1$
and different values of $e$. The dashed (dotted-dashed) curves correspond to
the periodic (non-periodic) even solutions. 
Again, both problems (periodic and non-periodic)
have a common solution corresponding to the full curve of Fig.
\ref{fig:sigma} in the limit of large $e$.
The monotonic decrease of $R_{\alpha}$ versus $s$ is in agreement with the stationary solutions of \cite{Avalos03} for the outer wall problem
and of \cite{Schubert01} for the inner wall problem.\\
In Fig. \ref{fig:mu} the
marginal curves $R_{\alpha}$ versus $\log_{10}(m)$ are given for $k=1$, $s=n=1$
and different values of $e$. The dashed (dotted-dashed) curves correspond to
the periodic (non-periodic) even solutions. 
Again, both problems (periodic and non-periodic)
have a common solution corresponding to the full curve of Fig.
\ref{fig:mu} in the limit of large $e$. This curve coincides with 
the full curve of Fig. \ref{fig:sigma}. Indeed for a given $k$ and in the limit of large $e$
we find from (\ref{MSdefinition}) and (\ref{pereven}) that $(M,S)=(m,s)/k$ for both problems. As (\ref{ap:carac})
is symmetric in $M$ and $S$, the full curves of Fig. \ref{fig:sigma} and Fig. \ref{fig:mu} coincide.
For the same reason the dashed curves corresponding to
 the periodic problem of Fig. \ref{fig:sigma} and Fig. \ref{fig:mu} coincide.
 Again the monotonic decrease of $R_{\alpha}$ versus $m$ is in agreement with the stationary solutions of \cite{Avalos03}\\
In Fig.
\ref{fig:perm} the isolines of $R_{\alpha}$ are sketched in the
($\log_{10}(s)$, $\log_{10}(m)$)-plane for $k=1$, $e=0.1$ and $n=m$. The isolines
of the periodic even solutions (dashed curves) are symmetric to
the straight line $m=s$ from the same symmetry arguments as above.
 This is not true for the
non-periodic even solutions (full lines). In that case the
asymmetry comes from the fact that $\sigma_3$ and $\mu_3$ do not play a
symmetric role as $\sigma_3=0$ and $\mu_3 \neq 0$.
In table \ref{tab:sm} we give the values of $R_{\alpha}$ for asymptotic values of $(s,m)$.
We note that they do not depend on the type of the problem either periodic or non periodic.\\
\begin{table}  
  \begin{center}   
  \begin{tabular}{|l|c|c|} \hline  
        & $s<<1$ & $s>>1$\\ \hline
 $m<<1$ &  3.5   & 2.04 \\ \hline
 $m>>1$ &  2.03   & 1 \\ \hline
\end{tabular}  
  \end{center}  
  \caption{Values of $R_{\alpha}$ for asymptotic values of $(s,m)$ and $k=1$, $e=0.1$, $n=m$
  and both problems periodic and non periodic.}   
\label{tab:sm}   
\end{table}
At that point we can make two conclusions. First in the non periodic case when varying the
electromagnetic properties of the outer wall, the roles of the conductivity and permeability are different.
In other words, a given jump of magnetic diffusivity $\eta=1/\sigma \mu$ between the fluid and the wall
can lead to different results depending on the respective jumps of the conductivity and permeability.
Second, for the periodic problem and $e=10^{-1}$, we see from Fig. \ref{fig:mu} that when increasing $m$ from $1$ to $10^3$
the reduction of the threshold is small ($4.5 \%$). Now we see from Fig. \ref{fig:perm} that such a reduction at $m=10^3$ can be compensated by the decrease of $s$ from 1 to 0.89. In other words the threshold is the same ($\sim$ 1.1) for $(s, m)=(1, 1)$ and $(s, m)=(0.89, 10^3)$.
Then the reduction of the threshold when using ferromagnetic steel inner walls (with a relative conductivity of order 1) instead of non ferromagnetic materials is negligible. Therefore it does not seem likely that the dynamo action in a FBR can be favored by the use of ferromagnetic assemblies at least in the framework of this simplified study. This is in disagreement with more elaborate models \cite{Soto99} which show that the jump of $m$ from 1 to $10^3$ reduces the threshold significantly enough to start the dynamo action inside the core of a FBR. A reason of discrepancy may come from the fact that in \cite{Soto99} $s$ is kept constant. Also it may come from the idealized geometry of the flow inside each assembly in \cite{Soto99} which is helical instead of being replaced by some $\alpha$-effect as here. Now let us consider the influence of a ferromagnetic belt surrounding the core of a FBR. From Fig. \ref{fig:mu} we see that, for the non periodic problem, from $m=1$ to $m=10^3$
the threshold reduction rate is significant (larger than $20\%$ for $e=10^{-1}$ and $e=1$). If now we assume that the ferromagnetic material is less conducting than the fluid then from Fig. \ref{fig:perm} jumping from $(s,m)=(1,1)$ to $s\le 1$ and $m=10^3$ leads to a minimum threshold reduction of $10\%$. Finally it is likely that an outer ferromagnetic belt can favour the dynamo action in the core of a FBR.

\subsection{Influence of the fluid permeability}
\begin{figure} 
  \begin{flushright} 
  \begin{tabular}{@{\hspace{0cm}}l@{\hspace{0cm}}c@{\hspace{0cm}}} 
    \raisebox{5cm}{$R_{\alpha}$} 
  \epsfig{file=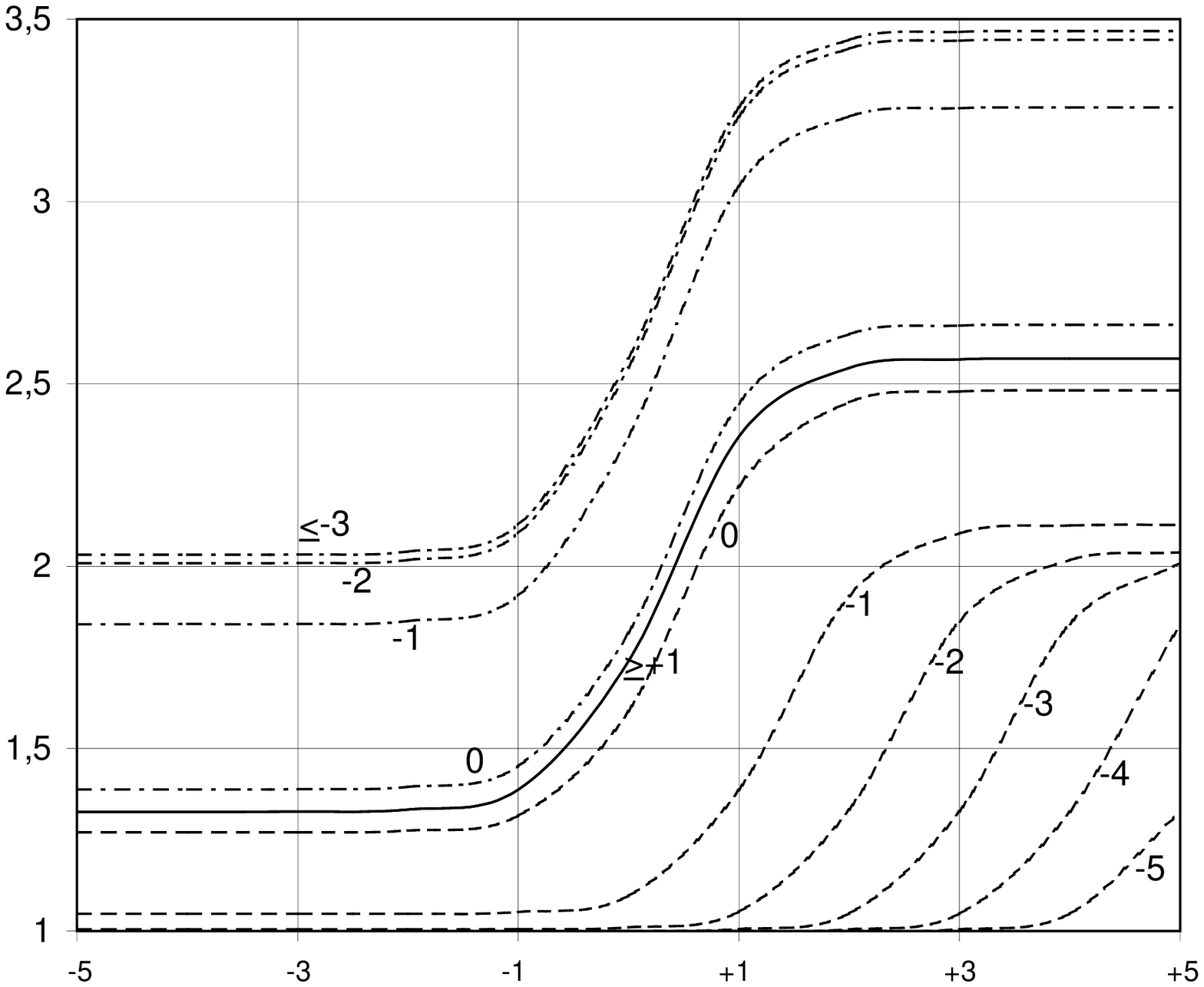,width=0.6\textwidth} 
    \\*[-1cm] 
     & \parbox{0.5\textwidth}{\hspace{-5cm} $\log_{10}(n/m)$} 
    \end{tabular} 
\end{flushright} 
    \caption{The marginal curves
$R_{\alpha}$ versus $\log_{10}(n/m)$ for $k=1$,
$s=n=1$ and different values of $e$. The labels indicate $\log_{10}(e)$.
The dashed (dotted-dashed) curves correspond to the periodic (non-periodic) even
solutions.}
    \label{fig:billes}
   \end{figure}

The influence of the fluid permeability can also be given by 
the expansions of table \ref{tab:asymptotics} for small values of both $k$ and
$ke$. We see indeed that $R_{\alpha}=O((n/m)^{1/2})$ for the non-periodic
even solutions and that $R_{\alpha}=O((1/m)^{1/2})$ in the limit of large
$1/m$  for the
periodic even solutions. Therefore for both problems a fluid
with large permeability leads to higher $R_{\alpha}$. This
is also confirmed by Fig.\ref{fig:billes} in which the
marginal curves $R_{\alpha}$ versus $n/m$
 are given for $k=1$, $s=n=1$ and different values of $e$. 
The dashed (dotted-dashed) curves correspond to the
periodic (non-periodic) even solutions.
Though $R_{\alpha}$ increases versus $n/m$ we must note however that $R_{\alpha}/\sigma_1 \mu_1$ decreases
versus $n/m$ (see also \cite{Avalos03}). This means that, keeping $R_{\alpha}$ constant and using a fluid of high permeability allows the experimenter to reduce significantly the flow intensity or the size of the device. Assuming as in \cite{Fauve03} that the power to drive an experiment is dissipated by turbulence, leads to a power proportional to $U^3$. Therefore using a fluid of high permeability would imply a significant reduction of the driving power.
This has also motivated some experimental studies \cite{Martin00}\cite{Frick02b}.

\subsection{Geometrical effects}
Following the same idea as in section \ref{sec:e}, we calculate the dissipation and work of the Lorentz force rates for each previous case (changing the conductivity or permeability of the wall or the fluid).  We find again that the change of ${\cal J}$ is always accompanied by a change of ${\cal S}$ and
that the dissipation in the wall is always negligible compared to the dissipation in the fluid. In
table \ref{tab:dishel}, we give some global informations on the behavior of ${\cal J}$ and ${\cal S}$ for the different previous cases.
For the same reasons as in section \ref{sec:e} we conclude that adding outer or inner walls with different conductivity or permeability has a pure geometrical effect on the field and current lines.
\begin{table}  
  \begin{center}   
  \begin{tabular}{|l c|c|c|c|c|} \hline  
        && $e$ $\nearrow$ & $s$ $\nearrow$& $m=n$ $\nearrow$&$1/m$ $\nearrow$\\ \hline
Non per. &$\cal S$& $\nearrow$ 7.7 & $\nearrow$ 1.17 & $\searrow$ 6.22  & $\nearrow$ 6.22 \\ \cline{2-6}
         &$\cal J$& $\searrow$ 6.76 & $\searrow$ 1.65 & $\searrow$ 11  & $\nearrow$ 11     \\ \cline{2-6}
$R_{\alpha}=$&${\cal J/S}$& $\searrow$ 52 & $\searrow$ 1.93 & $\searrow$ 1.77  & $\nearrow$ 1.77     \\ \hline 
Periodic &$\cal S$& $\searrow$ 2 & $\nearrow$ 1.14 & $\searrow$ 1.3 & $\nearrow$ 1.3 \\ \cline{2-6}
         &$\cal J$& $\nearrow$ 17 & $\searrow$ 1.77 & $\searrow$ 2.52  & $\nearrow$ 2.52     \\ \cline{2-6}
$R_{\alpha}=$ & ${\cal J/S}$& $\searrow$ 34 & $\searrow$ 2 & $\searrow$ 1.94  & $\nearrow$ 1.94     \\ \hline 
\end{tabular}  
  \end{center}  
  \caption{Global information on the evolution of $\cal S$ and $\cal J$ versus $e$ (for $k=0.01$) and versus $s$, $m=n$ and $1/m$ (for $k=1$).
The arrows indicate if the quantity grows or decays. The numbers indicate the factor of growth or decay.}   
\label{tab:dishel}   
\end{table}   
\subsection{Infinite conductivity or permeability}
Two usual ways to simplify the dynamo problem are to
solve the induction equation in region 1 only, with one of the following
boundary conditions at $x=1$:
\begin{equation}
\bB \cdot \bn = \bn \times \bE=0,
\label{eq:sinfty}
\end{equation}
or
\begin{equation}
\bn \times \bB = \bj \cdot \bn = 0.
\label{eq:minfty}
\end{equation}
The first boundary condition (\ref{eq:sinfty}) corresponds to region 2 being a perfect conductor whereas
the second boundary condition (\ref{eq:minfty}) corresponds to region 2 having an infinite permeability.
A priori we would expect to recover these two limits with our model taking the limit $s>>1$ or $m>>1$. Furthermore these limits should not depend on $e$ nor on the type of problem
periodic or non periodic which is considered. As can be seen from Fig. \ref{fig:sigma} and Fig. \ref{fig:mu} this is not true.
In fact it is known \cite{Landau69} that considering an ordinary body of electric conductivity $\sigma$ and magnetic permeability $\mu$, the
supraconductivity limit corresponds not only to $\sigma \rightarrow \infty$ but in addition to $\mu \rightarrow 0$ . Considering the double limit $s>>1$ and $m<<1$ in our model 
we indeed recover the boundary condition (\ref{eq:sinfty}).
A body of high permeability is also known to be a poor conductor and corresponds not only to $\mu \rightarrow \infty$ but in addition to $\sigma \rightarrow 0$ . Considering the double limit $m>>1$ and $s<<1$ in our model 
we indeed recover the boundary condition (\ref{eq:minfty}).
For both double limits the dynamo threshold is found to be $R_{\alpha} \approx 2$ as given in table \ref{tab:sm} \footnote{We believe that the fact that $R_{\alpha}$ is the same for both double limits is coincidental and probably related to our model.}.

\section{Conclusions}
\label{sec:conclusion}
For a stationary dynamo with either insulating or periodic boundary conditions, we showed that additional outer or inner walls change the geometry of the field and current lines in the fluid. This geometrical effect has an effect on the dynamo instability threshold. 
For given inner or outer walls, increasing their conductivity or permeability helps for dynamo action. In the other hand increasing the thickness
of the inner wall plays against the dynamo action (contrary to the outer wall for which increasing the thickness helps for dynamo action).
These conclusions are consistent with those obtained in other geometries for outer \cite{Avalos03}\cite{Ravelet05}\cite{Sarson96} as well as inner \cite{Schubert01} walls and we believe that they are generic in the sense that they do not depend on the generation process ($\alpha^2$ or else) as far as the solution stays stationary.
The detailed mechanism of the geometrical effect is however non trivial as the dissipation in the fluid is changed as well as the work of the Lorentz forces. In any case the dissipation in the walls is always negligible.\\
The usual boundary conditions used to describe a perfectly conducting or high permeability outer wall
are recovered with a double limit on $s$ and $m$, stressing that a supraconductor outer wall would correspond to $s \rightarrow \infty$ and $m \rightarrow 0$
and a high permeability outer wall to $m \rightarrow \infty$ and $s \rightarrow 0$.\\
As mentioned in section \ref{sec:alphaeffect}, an additional mean flow $\bmU$ like in the core of 
a fast breeder reactor might lead to oscillatory dynamo solutions. Then in this case as shown in \cite{Avalos03} for the Riga dynamo experiment, some additional
eddy currents in the walls may lead to significant dissipation in the walls reminiscent to a skin effect. Then the previous behavior of 
 $R_{\alpha}$ versus $e$, $s$ or $m$ would not be monotonic anymore. For the specific case of the core of a FBR, the conclusions of section \ref{sec:cp} should then be revised with respect to this additional mean flow $\bmU$. It is not clear a priori what would be the main effect of $\bmU$, either increasing or decreasing the dynamo threshold.
\\
\\
R. Avalos-Zu\~niga acknowledges the Mexican CONACYT for financial support.
\renewcommand{\theequation}{A.\arabic{equation}}
\setcounter{equation}{0}  
\section*{Appendix A. Proof \footnote{A similar proof of the principle of exchange of
stabilities has been given for an $\alpha^2$-dynamo with a constant $\alpha$ within an electrically conducting sphere
surrounded by an insulator \cite{Malkus75}.
} that $p$ has no imaginary part 
}  
Taking (\ref{eq:equations2})
 and the complex conjugate of
(\ref{eq:equations1}), we
easily find 
\begin{eqnarray}
\eta_l a_l\underline{a}_l''  -(\eta_l k^2+\underline{p}) {|a_l|}^2
-\alpha_l\underline{b}_la_l=0 \label{eq:equationsA1}\\ 
\eta_l b_l''\underline{b}_l  -(\eta_l k^2+p) {|b_l|}^2
-\alpha_lk^2\underline{b}_la_l=0 \label{eq:equationsB1}
\end{eqnarray}
where an underlined quantity means the complex conjugate of this quantity.
Integrating by part and using the boundary conditions (\ref{apBC}) we can show that
\begin{eqnarray}
\int_{x_{l-1}}^{x_l}a_l\underline{a}_l''dx&=&-\frac{\mu_l}{\mu_{l-1}}a_{l-1}\underline{a}_{l-1}'(x_{l-1})
+a_l\underline{a}_l'(x_l)\nonumber\\
&-&\int_{x_{l-1}}^{x_l}|a_l'|^2dx\label{eq:equationsC2}\\
\int_{x_{l-1}}^{x_l}b_l''\underline{b}_ldx&=&
-\frac{\eta_{l-1} \mu_l}{\eta_l \mu_{l-1}}b_{l-1}'\underline{b}_{l-1}(x_{l-1})
+b_l'\underline{b}_l(x_l)\nonumber\\
&-&\int_{x_{l-1}}^{x_l}|b_l'|^2dx
\end{eqnarray}
where $x_0=a_0=b_0=0$. For the non-periodic case $x_3=+\infty$ and
$a_3(x=+\infty)=b_3(x=+\infty)=0$. For the periodic case the periodicity implies
that $a_2\underline{a}_2'(x=R+e)=b'_2\underline{b}_2(x=R+e)=0$. The previous
relations do not depend on the parity of $a_1$ nor $b_1$. Combining the
integral of relations (\ref{eq:equationsA1}) and (\ref{eq:equationsB1}), we
find: \begin{eqnarray}
0=\sum_l\frac{1}{\mu_l}\int_{x_{l-1}}^{x_l}(&-&\eta_1 k^2 |a_l'|^2 + \eta_l
|b_l'|^2 - \eta_1 k^4 |a_l|^2\\ &+& \eta_l k^2 |b_l|^2
- \frac{\eta_1}{\eta_l}k^2 \underline{p} |a_l|^2 + p |b_l|^2)dx. \nonumber
\label{equationp}
\end{eqnarray}
Then it is straightforward to show that $\Im(p)=0$.

\renewcommand{\theequation}{B.\arabic{equation}}
\setcounter{equation}{0}  
\section*{Appendix B. Expression of the dissipation rate 
}  
Multiplying (\ref{inducmean2}) by $\bmB/\mu$  we obtain the following
 equation (where underlying means complex conjugate):
  \begin{equation}   
\frac{\partial }{\partial t}(\frac{|\bb|^2}{2\mu}) =
\frac{\partial}{\partial x}(\frac{\underline{\bb}\times \be}{\mu}) 
-\alpha(\bb-(\hat{\bz}.\bb)\hat{\bz})\cdot\underline{\bj}
-\frac{|\bj|^2}{\sigma}  \label{energ}  
\end{equation}  
with $\bb=(-ika,b,a')$, $\bj=\frac{1}{\mu}(-ikb,-\Delta a,b')$,
$\be=\bj/\sigma+\alpha(\bb-(\hat{\bz}.\bb)\hat{\bz})$. Applying the boundary
conditions (\ref{apBC}) we can show that 
$\int_0^{\Gamma} \frac{\partial}{\partial x}(\frac{\underline{\bb}\times
\be}{\mu})dx =0$ where $\Gamma=R+e\;\;(=+\infty)$ for the periodic (non-periodic) problem.
At the dynamo onset $\frac{\partial
}{\partial t}(|\bb|^2/2\mu) = 0$ and the Joule dissipation
$|\bj|^2/\sigma$ is equal to the alpha-dynamo source
$-\alpha(\bb-(\hat{\bz}.\bb)\hat{\bz})\cdot\underline{\bj}$.
In that case (at the onset) the expressions of
the dimensionless dissipation $\cal{J}$ and work of the Lorentz forces $\cal{S}$ rates
are given by
\begin{equation}  
{\cal J}=
\frac{\int_0^{1+\hat{e}}\frac{\hat{\eta}}{\hat{\mu}}((\hat{k}^2+\frac{R_{\alpha}^2}{\hat{\eta}^2})|\hat{b}|^2+|\hat{b}'|^2)d\hat{x}}
{\int_0^{\hat{\Gamma}}\frac{1}{2\hat{\mu}}(\hat{k}^2|a|^2+|\hat{b}|^2+|a'|^2)d\hat{x}},
\end{equation}  
\begin{equation}  
{\cal S}=
\frac{\int_0^{1}\frac{1}{\hat{\mu}}(\hat{b}a''-2k^2a \hat{b})d\hat{x}}
{\int_0^{\hat{\Gamma}}\frac{1}{2\hat{\mu}}(\hat{k}^2|a|^2+|\hat{b}|^2+|a'|^2)d\hat{x}},
\end{equation} 
where $\hat{x}=x/R$, $\hat{\eta}=\eta/\eta_1$,
$\hat{\mu}=\mu/\mu_1$, $\hat{k}=Rk$ and $\hat{b}=Rb$.

%
%

\end{document}